\documentstyle[12pt]{article}
\makeatletter
\def\wgas@ver{2.2}
\def\wgas@pageid{\xdef\@thefnmark{\null}
\@footnotetext{This manuscript was prepared with the
		   AAS WGAS \LaTeX\ macros v\wgas@ver}}
\ifnum\@ptsize<2
\fi
\ps@plain
\def\@tightleading{1.1}
\def\@doubleleading{1.6}
\def\baselinestretch{\@doubleleading}
\def\tightenlines{\def\baselinestretch{\@tightleading}}
\def\received#1{\gdef\@recvdate{#1}} \received{\relax}
\def\accepted#1{\gdef\@accptdate{#1}} \accepted{\relax}
\def\journalid#1#2{\gdef\@jourvol{#1}\gdef\@jourdate{#2}}
\def\articleid#1#2{\gdef\@startpage{#1}\gdef\@finishpage{#2}}
\def\@rcvaccrule{\vrule\@width1.75in\@height0.5pt\@depth0pt}
\def\dates{{\center\small{\it Received:}\space%
\if\@recvdate\relax\@rcvaccrule\else\@recvdate\fi;%
\hspace{1.5em}{\it Accepted:}\space%
\if\@accptdate\relax\@rcvaccrule\else\@accptdate\fi%
\endcenter}}
\let\ltx@abstract=\abstract
\def\abstract{\dates\ltx@abstract}
\def\title#1{{\def\baselinestretch{\@tightleading}
\center\large\bf#1\endcenter}}
\def\author#1{{\topsep 0pt\center\normalsize#1\endcenter}}

\def\altaffilmark#1{$^{#1}$}
\def\altaffiltext#1#2{\footnotetext[#1]{#2}}
\skip\footins 4ex plus 1ex minus .5ex
\long\def\@footnotetext#1{\insert\footins{
\footnotesize
\interlinepenalty\interfootnotelinepenalty
\splittopskip\footnotesep
\splitmaxdepth \dp\strutbox \floatingpenalty \@MM
\hsize\columnwidth \@parboxrestore
\edef\@currentlabel{\csname p@footnote\endcsname\@thefnmark}\@makefntext
{\rule{\z@}{\footnotesep}\ignorespaces
#1\strut}}}
\long\def\@makefntext#1{\noindent\hbox to\z@{\hss$^{\@thefnmark}$}#1}

\def\tablenotetext#1#2{
\@temptokena={\vspace{.5ex}{\noindent\llap{$^{#1}$}#2}\par}
\@temptokenb=\expandafter{\tblnote@list}
\xdef\tblnote@list{\the\@temptokenb\the\@temptokena}}
\def\spewtablenotes{
\ifx\tblnote@list\@empty
\else
\let\@temptokena=\tblnote@list
\gdef\tblnote@list{\@empty}
\vspace{4.5ex}
\footnoterule
\vspace{.5ex}
{\footnotesize\@temptokena}
\fi}
\newtoks\@temptokenb
\def\tblnote@list{}
\def\endtable{\spewtablenotes\end@float}
\@namedef{endtable*}{\spewtablenotes\end@dblfloat}

\def\@xfloat#1[#2]{\ifhmode \@bsphack\@floatpenalty -\@Mii\else
\@floatpenalty-\@Miii\fi\def\@captype{#1}\ifinner
\@parmoderr\@floatpenalty\z@
\else\@next\@currbox\@freelist{\@tempcnta\csname ftype@#1\endcsname
\multiply\@tempcnta\@xxxii\advance\@tempcnta\sixt@@n
\@tfor \@tempa :=#2\do
{\if\@tempa h\advance\@tempcnta \@ne\fi
\if\@tempa t\advance\@tempcnta \tw@\fi
\if\@tempa b\advance\@tempcnta 4\relax\fi
\if\@tempa p\advance\@tempcnta 8\relax\fi
}\global\count\@currbox\@tempcnta}\@fltovf\fi
\global\setbox\@currbox\vbox\bgroup
\@normalsize
\boxmaxdepth\z@
\hsize\columnwidth \@parboxrestore}
\def\@keywordtext{Subject headings}
\def\@keyworddelim{---}
\def\keywords#1{\vspace*{-.7ex}
\if@twocolumn\noindent{\small{\it\@keywordtext:\/}\space\@kwds{#1}}
\else{\quote\small{\it\@keywordtext:\/}\space\@kwds{#1}\endquote}
\fi}

\def\@kwds#1{\def\@kwddlm{}\@for\@kwd:=#1\do
{\@kwddlm\def\@kwddlm{\space\@keyworddelim\penalty\@m\space}{\@kwd}}}
\def\section{\@startsection {section}{1}{\z@}{2.3ex plus 1ex minus
.2ex}{1.5ex plus .2ex}{\normalsize\bf}}
\def\subsection{\@startsection{subsection}{2}{\z@}{2ex plus 1ex minus
.2ex}{1ex plus .2ex}{\normalsize\bf}}
\def\subsubsection{\@startsection{subsubsection}{3}{\z@}{2ex plus
1ex minus .2ex}{1ex plus .2ex}{\normalsize\it}}
\def\acknowledgments{\vskip 3ex plus .8ex minus .4ex}

\def\mathwithsecnums{
\@newctr{equation}[section]
\def\theequation{\hbox{\normalsize\arabic{section}-\arabic{equation}}}}
\def\references{\subsection*{REFERENCES}
\bgroup\parindent=0pt\parskip=\itemsep
\def\refpar{\par\hangindent=3em\hangafter=1}}
\def\endreferences{\refpar\egroup\wgas@pageid}

\def\endthebibliography{\endlist\wgas@pageid}
\def\@biblabel#1{\relax}
\def\@cite#1#2{(#1\if@tempswa , #2\fi)}

\def\@citex[#1]#2{\if@filesw\immediate\write\@auxout{\string\citation{#2}}\fi
\def\@citea{}\@cite{\@for\@citeb:=#2\do
{\@citea\def\@citea{;\penalty\@m\ }\@ifundefined
{b@\@citeb}{\@warning
{Citation `\@citeb' on page \thepage \space undefined}}%
{\csname b@\@citeb\endcsname}}}{#1}}
\let\jnl@style=\rm
\def\ref@jnl#1{{\jnl@style#1}}
\def\aj{\ref@jnl{AJ}}
\def\araa{\ref@jnl{ARA\&A}}
\def\apj{\ref@jnl{ApJ}}
\def\apjl{\ref@jnl{ApJ}}
\def\apjs{\ref@jnl{ApJS}}
\def\applopt{\ref@jnl{Appl.Optics}}
\def\apss{\ref@jnl{Ap\&SS}}
\def\aap{\ref@jnl{A\&A}}
\def\aapr{\ref@jnl{A\&A~Rev.}}
\def\aaps{\ref@jnl{A\&AS}}
\def\azh{\ref@jnl{AZh}}
\def\baas{\ref@jnl{BAAS}}
\def\jrasc{\ref@jnl{JRASC}}
\def\memras{\ref@jnl{MmRAS}}
\def\mnras{\ref@jnl{MNRAS}}
\def\pra{\ref@jnl{Phys.Rev.A}}
\def\prb{\ref@jnl{Phys.Rev.B}}
\def\prc{\ref@jnl{Phys.Rev.C}}
\def\prd{\ref@jnl{Phys.Rev.D}}
\def\prl{\ref@jnl{Phys.Rev.Lett}}
\def\pasp{\ref@jnl{PASP}}
\def\pasj{\ref@jnl{PASJ}}
\def\qjras{\ref@jnl{QJRAS}}
\def\skytel{\ref@jnl{S\&T}}
\def\solphys{\ref@jnl{Solar~Phys.}}
\def\sovast{\ref@jnl{Soviet~Ast.}}
\def\ssr{\ref@jnl{Space~Sci.Rev.}}
\def\zap{\ref@jnl{ZAp}}

\let\apjlett=\apjl

\def\la{\mathrel{\hbox{\rlap{\hbox{\lower4pt\hbox{$\sim$}}}\hbox{$<$}}}}
\def\ga{\mathrel{\hbox{\rlap{\hbox{\lower4pt\hbox{$\sim$}}}\hbox{$>$}}}}

\def\fs{\hbox{$.\!\!^{\rm s}$}}
\def\fdg{\hbox{$.\!\!^\circ$}}

\def\farcs{\hbox{$.\!\!^{\prime\prime}$}}

\newcount\lecurrentfam
\def\LaTeX{\lecurrentfam=\the\fam \leavevmode L\raise.42ex
\hbox{$\fam\lecurrentfam\scriptstyle\kern-.3em A$}\kern-.15em\TeX}
\def\foot1#1{
\footnotemark
\addtocounter{footnote}{-1}
{\hrule
\vspace{\baselineskip}
\footnotemark
#1
\vspace{\baselineskip}
\hrule
}
}
\makeatother
\newcommand{\Msun} {M_\odot}
\newcommand{\etal} {{\it et~al.\ }}
\tightenlines
\received{1992 October 14}
\accepted{1992 November 17}

\begin{document}
\title{The Masses of Two Binary Neutron Star Systems}
\author{S. E. Thorsett,\altaffilmark{1}
 Z. Arzoumanian,\altaffilmark{2}
 M. M. McKinnon,\altaffilmark{3}
 J. H. Taylor\altaffilmark{2}}

\altaffiltext{1}{Owens Valley Radio Observatory, 105--24,
  California Institute of Technology, Pasadena, CA 91125}
\altaffiltext{2}{Joseph Henry Laboratories and Department of
  Physics,
  Princeton University, Princeton, NJ 08544}
\altaffiltext{3}{National Radio Astronomy Observatory,
  P.O. Box 2,
  Green Bank, WV 24944}

\begin{abstract}
The measurement or constraint of the masses of neutron stars and
their binary companions tests theories of neutron star structure
and of pulsar formation and evolution.  We have measured the rate
of the general relativistic advance of the longitude of
periastron for the pulsar PSR B1802$-$07:
$\dot\omega=0\fdg060\pm0\fdg009\,\mbox{yr}^{-1}$, which implies a
total system mass, pulsar plus companion star, of
$M=1.7\pm0.4\,\Msun$.  We also present a much improved
measurement of the rate of periastron advance for PSR B2303+46:
$\dot\omega = 0\fdg0099\pm0\fdg0002\,\mbox{yr}^{-1}$, implying
$M=2.53\pm0.08\,\Msun$ for this system.  We discuss the available
constraints on distribution of mass between the pulsars and their
companions, and we compare the pulsar masses with other
determinations of neutron star masses.
\end{abstract}

\section{Introduction}

Since the pioneering calculations of Oppenheimer and Volkoff
(1939)\nocite{ov39}, physical models have predicted a limited
range of neutron star masses.  Nearly all modern equations of
state for nuclear matter in bulk require
$0.1\Msun\la m_{\rm ns} \la 3\Msun$, assuming that general
relativity is valid in the strong field regime (see reviews by
Wheeler 1966, Hartle 1978, Baym \& Pethick 1979, Shapiro \&
Teukolsky 1983;\nocite{whe66,har78,bp79,st83} see also
Bahcall, Lynn, \& Selipsky 1990).\nocite{bls90}  Much tighter
limits on the masses of radio and x-ray pulsars are suggested by
the origin of these neutron stars in collapses of the degenerate
cores of highly evolved massive stars.  If collapse occurs when
accretion of the stellar envelope onto the degenerate core
increases the mass of the core to about the Chandrasekhar mass,
$m_{\rm c}\approx1.4\Msun$, then the gravitational mass (baryonic
mass minus binding energy) of the resulting neutron star should
be $m_{\rm ns}\la1.4\Msun$.  Detailed models generally produce
neutron stars with $m_{\rm ns}\ga1.15\Msun$, with typical values
around $1.3\Msun$ \cite{woo87}.

Spectroscopic observations of seven high-mass x-ray binary
systems have allowed determination of the masses of their neutron
star components to precisions of between 10 and 50\% (for reviews, see
Joss \& Rappaport 1976\nocite{jr76}, Bahcall 1978\nocite{bah78},
Nagase 1989\nocite{nag89}).  The discovery of the first radio
pulsar in a relativistic binary orbit \cite{ht75a} has permitted
mass determinations of the neutron stars comprising the PSR B1913+16
system to better than 0.1\% precision
\cite{tw89,tay92a}.  With more than twenty radio pulsars now
known in binary systems, including four or five with neutron star
companions, many more neutron star masses have become potentially
measurable.  In this {\em Letter} we present the latest results
of two such measurements, for the binary pulsar systems PSR
B1802$-$07 and PSR B2303+46.  For each system we have measured
the orbital period, eccentricity, and projected semi-major axis,
as well as the rate of advance of the longitude of periastron.
The results provide unambiguous measurements of the total masses
of the two binary systems, as well as useful constraints on the
individual component masses.

\section{Observations}

We have observed both PSR B1802$-$07 and PSR B2303+46 at the Very Large
Array in Socorro, New Mexico, and at the 140~ft telescope of the
National Radio Astronomy Observatory at Green Bank, West Virginia.

The pulsar PSR B1802$-$07 is in the globular cluster NGC~6539
\cite{dlb+90a}.
All observations of this pulsar at the Very Large Array were made
using the Princeton Mark~III pulsar timing system \cite{skn+92}, on
19~days between March 1991 and September 1992, as part of a general
study of binary and millisecond pulsars \cite{tho91b}.
The VLA was used as a phased array, with a bandpass centered at
1665~MHz divided by a filter bank into fourteen adjacent 4~MHz
channels in each circular polarization.  Because of other
filtering by VLA electronics, the effective total bandwidth was
limited to 46~MHz.  Detected signals in each channel were
averaged synchronously with the pulsar period for intervals of
2--5 minutes, and the start time of each integration was recorded
from a local clock traceable to the best atomic time scales by
means of a Global Positioning System (GPS) receiver.
In addition, 14 days of observations at the Green Bank 140~ft telescope
were carried out, with equipment described below,
between January and October 1992, at frequencies near 800 and 1330~MHz.
Average pulse profiles were recorded at regular intervals at both sites,
and subsequent
processing included removal of differential time delays between
channels (caused by dispersion in the ionized interstellar
medium) and reduction of each integrated profile to a single
pulse time-of-arrival, using standard techniques \cite{tay90a}.

PSR B2303+46 was discovered by Dewey {\it et al.}~(1985)
\nocite{dtws85} and quickly identified as a member of a binary
system with high orbital eccentricity \cite{std85}.
At the VLA, we used the Mark~III timing system on
three days.  The center frequency for these observations was
335~MHz, the channel bandwidth was 2~MHz, and total bandwidth was
20~MHz. At Green Bank, we made observations on 32 days between August
1989 and October 1992, using a
digital Fourier transform spectrometer to divide a 40~MHz
bandpass centered near 400 or 800~MHz into 512 frequency
channels, which were detected, averaged, and reduced in a manner
similar to the Mark~III timing data.  In addition to these new
observations of PSR B2303+46, our analysis also included the
pulse arrival times recorded on 58~days between 1985 and 1987,
previously described by Taylor and Dewey (1988)\nocite{td88}.

\section{Analysis}

The pulse times of arrival were analyzed using standard
techniques \cite[and references therein]{tw89}.  The {\sc tempo}
pulsar timing software package was used to reduce the topocentric
arrival times to the reference frame of the solar system
barycenter, and then to fit a model of each pulsar's spin,
astrometric, and orbital elements to the data.  The resulting
parameters and their uncertainties are listed in
Table~\ref{tab:params}.  They include the celestial coordinates,
the pulsar period and spin-down rate at a stated epoch, the
dispersion measure, and the orbital period $P_b$, eccentricity
$e$, projected semi-major axis $x\equiv a_1\sin i/c$, time of
periastron passage $T_0$, longitude of periastron $\omega$, and
rate of advance of periastron $\dot\omega$.  (Here $i$ is the
angle between the line of sight toward the pulsar and the
right-handed normal to the orbit, and $c$ is the speed of
light.)  Finally, an adjustable time offset was introduced in the
multi-parameter fit to allow for differing instrumental delays at the
two observatories.
For both pulsars the five ``Keplerian'' orbital
parameters have been measured with accuracies of 5 or more
significant digits.  The ``post-Keplerian'' parameter
$\dot\omega$ is measured with less accuracy, approximately 15\%
for PSR B1802$-$07 and 2\% for PSR B2303+46.  The uncertainties
quoted in Table~\ref{tab:params} include both statistical effects
and our best estimates of possible systematic errors, and are
intended to be realistic 68\% confidence limits.

The orbital period and semi-major axis of any binary pulsar
system define the pulsar mass function,
\begin{equation}
\label{eq:f1}
f_1\equiv\frac{(m_2\sin i)^3}{(m_1+m_2)^2} =
  \frac{x^3}{(P_b/2\pi)^2} \left(\frac{1}{T_\odot}\right)
  ~M_\odot\,,
\end{equation}
where $m_1$ and $m_2$ are the pulsar and companion mass in solar
units, $T_\odot\equiv GM_\odot/c^3=4.925490947\times10^{-6}\,$s,
and $G$ is the Newtonian constant of gravity.  For PSRs
B1807$-$02 and 2303+46 an additional mass constraint is provided
by our measurement of $\dot\omega$.  In both of these binary
systems we expect non-relativistic contributions to $\dot\omega$,
such as those caused by tidal or rotational distortions of the
stars, to be completely negligible;  consequently, the general
relativistic expression for $\dot\omega$ yields a solution for
the total system mass,
\begin{equation}
\label{eq:M}
M \equiv m_1 + m_2 = \frac{1}{3\sqrt{3}}
  \left(\frac{P_b}{2\pi}\right)^{5/2}
  \left(1-e^2\right)^{3/2} \dot\omega^{3/2}
  \left(\frac{1}{T_\odot}\right)~M_\odot\,.
\end{equation}

In principle, the measurement of one or more additional
parameters involving $m_1$, $m_2$, and $\sin i$ would allow an
unambiguous solution for all three quantities.  Potential
candidates for additional observables include further
relativistic effects (e.g., time dilation and gravitational
redshift, orbital period derivative, and ``Shapiro delay'');
measurements of the interstellar scintillation timescale to yield
the transverse orbital velocity of the pulsar, and hence the
orbital inclination \cite{lyn84}; and, for PSR B2303+46, the
discovery of pulsations from its neutron star companion, whose
orbital motion would provide an accurate value for the mass
ratio.  Unfortunately, for these two binary pulsars the prospects
for any of these measurements appear remote.  Expected values of
the additional post-Keplerian parameters are well below present
levels of detectability, even for observations extending over
several decades (see Damour \& Taylor 1992, Taylor 1992, for
further details).~\nocite{dt92,tay92a}  Moreover, both pulsars
are too faint for the necessary scintillation measurements to be
feasible with current generation telescopes, and attempts to
detect the companion of PSR B2303+46 as a radio pulsar have so
far been unsuccessful.

We are left, therefore, with total mass determinations whose
uncertainties are dominated by the uncertainties in $\dot\omega$,
plus firm constraints linking the values of $m_1$, $m_2$, and
$\sin i$ through the well-determined mass functions.  The
resulting measurements and limits are displayed for the two
pulsars in Figures~1 and 2. In these illustrations the sloping
straight lines flanked by dashed lines delimit 68\% confidence
regions for the total mass, $M\equiv m_1+m_2$, of each system.
The solid curves rising from left to right correspond to $\sin
i=1$, and the requirement $|\sin i|\le1$ excludes any mass
combinations below these limits.  Values of $m_1$ and $m_2$
anywhere within the unshaded regions are permitted by our data;
however, in a probabilistic sense (for random orientations of the
pulsar orbital planes) there is a 68\% chance that
$\cos i>0.68$, and therefore that the correct component masses
will be found below the dashed curves corresponding to
$\cos i=0.68$.  At the bottom of Table~\ref{tab:params}
we list mass limits corresponding to the relevant corners of the
68\% confidence regions in Figures~1 and 2.  The results are
fully consistent with our expectations that the companion of PSR
B1802$-$07 is a white dwarf, while the companion of PSR B2303+46
is most likely another neutron star.

\section{Discussion}

The orbital period of PSR B1802$-$07 is similar to those of
several other binary pulsars with low mass ($\la0.7\Msun$)
companions, including PSRs B0021$-$72E, B0655+64, B1639+36B,
B1831$-$00, and B1855+09.  However, its orbital eccentricity is
at least an order of magnitude larger than any of the others.
This fact is almost certainly related to peculiarities in the
evolution of this system.  Rasio and Shapiro (1991)\nocite{rs91}
have described a model of its formation in which the neutron star
collides (in the dense globular cluster environment of NGC 6539)
with a $\sim 0.8\Msun$~giant.  Hydrodynamical calculations showed
that the giant's envelope would be ejected within a few orbits,
leaving a helium core in an eccentric orbit around the pulsar,
together with a massive accretion disk.  During a low-mass x-ray
binary phase, the neutron star would accrete from this disk, and
the orbital eccentricity would decrease.  The disk is disrupted
too soon for complete circularization of the orbit or for
neutron-star spin-up to millisecond periods.  This model is
entirely consistent with our mass determinations.  However, it is
also possible that a more common evolutionary path produced a
nearly circular orbit that was subsequently made more eccentric
by one or more encounters with other cluster stars.

Our results for PSR B2303+46 are consistent with those of Taylor
\& Dewey (1988) and Lyne \& Bailes (1990),\nocite{td88,lb90} and
have considerably higher accuracy.  There is every reason to
believe that, as suggested previously, the companion of this
pulsar is another neutron star.  The companion mass is at least
$1.15\Msun$, and if we assume a pulsar mass of at least
$1.2\Msun$, then $m_2<1.33\Msun$.  The {\it average} mass of the
two neutron stars in this system is very well constrained:
$m_{\rm ave}=1.27\pm0.04$.  More sensitive searches for pulsed
emission from the companion would probably be worthwhile.

In Figure~3 we display our new mass constraints for three neutron
stars, together with mass measurements for 14 other neutron
stars.  All of these mass measurements are consistent with
$m_{\rm ns}=1.35\pm0.27\Msun$, indicated by the vertical dashed
lines in Figure~3.  Indeed, all measurements except those for
Her~X-1 and Vela~X-1 are consistent with a considerably narrower
range, $1.35\pm0.10\Msun$.  The weight of observational evidence
appears to say that although the nuclear equation of state may
permit stable neutron stars with masses down to $\sim0.1M_\odot$,
nature prefers to form them in a narrow mass range just below the
Chandrasekhar limit.

\acknowledgments
We thank D.~Frail, A.~Fruchter, M.~Goss, T.~Hankins, S.~Kulkarni,
R.~Manchester, and D.~Nice for valuable discussions and other
assistance.  The 140~ft telescope and the VLA are facilities of
the National Radio Astronomy Observatory, operated by Associated
Universities, Inc., under cooperative agreement with the National
Science Foundation.   Our work was supported in part by grants
from the NSF to Princeton and Caltech, and S.E.T.~is a Robert
A.~Millikan Research Fellow in Physics.

\clearpage
\begin{table}
\begin{center}
\caption{\label{tab:params} Measured parameters of the two binary
 pulsar systems.$^a$}
\begin{tabular}{lll}
\hline
Parameter & \multicolumn{1}{c}{PSR B1802$-$07} &
  \multicolumn{1}{c}{PSR B2303+46} \\
\hline
Right ascension (J2000) \dotfill & $18^{\rm h}\,04^{\rm
m}\,49\fs896(2)$ &
  $23^{\rm h}\,05^{\rm m}\,55\fs842(17)$ \\
Declination (J2000) \dotfill  & $-07^\circ\,35'\,24\farcs65(11)$
&
$+47^{\circ}\,07'\,45\farcs32(17)$ \\
Right ascension (B1950) \dotfill  & $18^{\rm h}\,02^{\rm
m}\,07\fs213(2)$ &
  $23^{\rm h}\,03^{\rm m}\,39\fs180(17)$ \\
Declination (B1950) \dotfill  &
$-07^{\circ}\,35'\,39\farcs57(11)$ &
  $+46^{\circ}\,51'\,31\farcs87(17)$ \\
Period (ms) \dotfill  & 23.10085521162(3) & 1066.371071565(16) \\
Period derivative ($10^{-19}$) \dotfill  & 4.75(9) & 5690.9(1.6)
  \\
Epoch (MJD) \dotfill  & 48540.0 & 46107.0 \\
Dispersion measure (cm$^{-3}\,$pc) \dotfill  & 186.38(3) &
 62.06(3) \\
Orbital period, $P_b$ (s) \dotfill  & 226088.36(4) &
1066136.648(15) \\
Projected semi-major axis, $x$ (lt-s) \dotfill  & 3.92047(4) &
  32.6878(3) \\
Eccentricity, $e$ \dotfill  & 0.211999(15) & 0.658369(9) \\
Longitude of periastron, $\omega$ (deg) \dotfill  & 164.928(18) &
35.0776(7) \\
Time of periastron passage, $T_0$ (MJD) \dots & 49401.19075(12) &
  47452.560747(17) \\
Advance of periastron, $\dot\omega$ (deg yr$^{-1}$) \dotfill &
0.060(9) &
  0.0099(2) \\
\\
Total mass, $M~(M_\odot)$ \dotfill & $1.7\pm0.4$ &
  $2.53\pm0.08$ \\
Pulsar mass, $m_1~(M_\odot)$ \dotfill & $1.4^{+0.4}_{-0.3}$ &
  $1.16\pm0.28$ \\
Companion mass, $m_2~(M_\odot)$ \dotfill & $0.33^{+0.13}_{-0.10}$
  & $1.37\pm0.24$ \\
\hline
\end{tabular}
\end{center}
\noindent $^a$Figures in parentheses are uncertainties in the
last digit quoted; all uncertainties are $1\sigma$ estimates,
inclusive of possible systematic effects.  J2000 coordinates are
based on the DE200 ephemeris of the Jet Propulsion Laboratory,
while B1950 coordinates refer to the Center for Astrophysics
PEP740R ephemeris.
\end{table}

\clearpage
\begin{figure}
\caption{\label{m1m2_1802}  The $m_1$-$m_2$ plane for PSR
B1802$-$07.  Hatched regions are excluded by limits on the total
system mass, the requirement that $|\sin i|\le1$, or both.
Dashed curve corresponds to $\cos i=0.68$, and marks the upper boundary
of a region with 68\% probability of containing the correct masses,
as described in the text.}
\end{figure}

\begin{figure}
\caption{\label{m1m2_2303}
The $m_1$-$m_2$ plane for PSR B2303+46. See caption for
Figure~1.}
\end{figure}

\begin{figure}
\caption[masses]{\label{masses}
Measured masses of 17 neutron stars.  Objects in massive x-ray
binaries are at the top, radio pulsars and their companions at
bottom. The lower mass limits for PSRs B1802$-$07 and B2303+46
and upper mass limit for the companion of PSR B2303+46 assume
$\cos i<0.68$, and are probabilistic in nature (see text). Dashed
vertical lines enclose a $\pm20$\% mass range around $1.35\Msun$.
References:
PSRs B1802$-$07, B2303+46: this work;
PSR B1855+09: Ryba and Taylor (1991)\nocite{rt91a};
PSR B1534+12: Wolszczan (1991)\nocite{wol91a};
PSR B1913+16: Taylor and Weisberg (1989)\nocite{tw89};
PSR B2127+11C: Anderson {\it et al.} (1992, preprint);
4U~1538-52: Reynolds, Bell, and Hilditch (1992)\nocite{rbh92};
4U~1700-37: Heap and Corcoran (1992)\nocite{hc92};
other x-ray pulsars: Nagase (1989)\nocite{nag89}.}
\end{figure}
\end{document}